\documentclass[sigconf, nonacm]{acmart}

\newcommand\vldbdoi{paperX.pdf}
\newcommand\vldbyear{2021}
\newcommand\vldbauthors{\authors}
\newcommand\vldbtitle{\shorttitle} 
\newcommand\vldbavailabilityurl{}
\newcommand\vldbpagestyle{plain} 

\begin{document}
\title{Finding NeMo: Fishing in banking networks using network motifs}

\author{Xavier Fontes }
\affiliation{%
  \institution{Feedzai}
  \city{Porto}
  \state{Portugal}
}
\email{xavier.fontes@feedzai.com}

\author{David Aparício}
\affiliation{%
  \institution{Feedzai}
  \city{Porto}
  \country{Portugal}
}
\email{david.aparicio@feedzai.com}

\author{Maria In\^es Silva}
\affiliation{%
  \institution{Feedzai}
  \city{Lisbon}
  \country{Portugal}
}
\email{ines.silva@feedzai.com}

\author{Beatriz Malveiro}
\affiliation{%
  \institution{Feedzai}
  \city{Lisbon}
  \country{Portugal}
}
\email{beatriz.malveiro@feedzai.com}

\author{Jo\~ao Tiago Ascens\~ao}
\affiliation{%
  \institution{Feedzai}
  \city{Lisbon}
  \country{Portugal}
}
\email{joao.ascensao@feedzai.com}

\author{Pedro Bizarro}
\affiliation{%
  \institution{Feedzai}
  \city{Lisbon}
  \country{Portugal}
}
\email{pedro.bizarro@feedzai.com}

\begin{abstract}
Banking fraud causes billion-dollar losses for banks worldwide. In fraud detection, graphs help understand complex transaction patterns and discovering new fraud schemes. This work explores graph patterns in a real-world transaction dataset by extracting and analyzing its network motifs. Since banking graphs are heterogeneous, we focus on heterogeneous network motifs. Additionally, we propose a novel network randomization process that generates valid banking graphs. From our exploratory analysis, we conclude that network motifs extract insightful and interpretable patterns.
\end{abstract}

\maketitle

\pagestyle{\vldbpagestyle}
\begingroup\small\noindent\raggedright\textbf{Reference Format:}\\
\vldbauthors. \vldbtitle. In the 2nd Workshop on Search, Exploration, and Analysis in Heterogeneous Datastores (SEA Data \vldbyear).\\
\endgroup
\begingroup
\renewcommand\thefootnote{}\footnote{\noindent
Copyright {\textcopyright} {\vldbyear} for the individual papers by the papers' authors. Copyright {\textcopyright} {\vldbyear}  for the volume as a collection by its editors. This volume and its papers are published under the Creative Commons License Attribution 4.0 International (CC BY 4.0).\\
Published in the Proceedings of the 2nd Workshop on Search, Exploration, and Analysis in Heterogeneous Datastores, co-located with VLDB {\vldbyear} (August 16-20, {\vldbyear}, Copenhagen, Denmark) on CEUR-WS.org.
}\addtocounter{footnote}{-1}\endgroup

\ifdefempty{\vldbavailabilityurl}{}{
\vspace{.3cm}
\begingroup\small\noindent\raggedright\textbf{PVLDB Artifact Availability:}\\
The source code, data, and/or other artifacts have been made available at \url{\vldbavailabilityurl}.
\endgroup
}

\section{Introduction}

Payment information theft renders online transactions susceptible to fraud. Once detected, fraud entails a reimbursement from the cardholder's bank, leading to monetary costs to financial institutions and customer friction.

Fraud detection thus requires a deep understanding of the underlying fraud patterns, and graphs offer an intuitive way to visualize these patterns. One can further leverage graph mining in transaction data to understand fraud schemes in a wide range of applications. \citet{hajdu2020temporal} developed a methodology to identify cycles in transaction networks as a means to detect fraudulent expenses. \citet{micale2019fast} retrieved the most frequent patterns in a relationship network of people involved in the Panama papers to identify the most relevant money laundering structures.

In this work, we build networks from a real-world banking dataset of card purchases and apply a widely used graph mining tool -- \emph{heterogeneous network motifs}~\cite{ribeiro2014discovering, rossi2019heterogeneous}. Analyzing recurring patterns in real banking networks sets a foundation for understanding how fraud materializes in transaction data. From there, one can extract insights about how legitimate and fraudulent transactions "behave" and aid both fraud detection systems and fraud analysts.

To the best of our knowledge, this work is the first to find and explore heterogeneous network motifs in a real-world banking setting.  Notably, we have two main contributions from this work:

\begin{itemize}
	\item We propose a randomization process (i.e., a null model) adequate to banking datasets, a vital component for the definition of network motifs used to provide the baseline frequencies of each subgraph (detailed in Section~\ref{method_network_motifs_null_model}).
	
	\item We extract network motifs from graphs built from card transaction data and review them thoroughly. We include an analysis of how the motif significance evolves as more random networks are used (detailed in Section~\ref{sec:results}).
\end{itemize}

The remainder of this paper is organized as follows. Section~\ref{sec:method} presents our method and discusses the key components necessary for our analysis. We describe the data and present results in Section~\ref{sec:results}. Usage scenarios are proposed in Section~\ref{sec:usage}. We put forward our main takeaways and offer directions for future work in Section~\ref{sec:conclusion}.

\begin{figure*}[!t]
	\centering
	\includegraphics[width=0.9\textwidth]{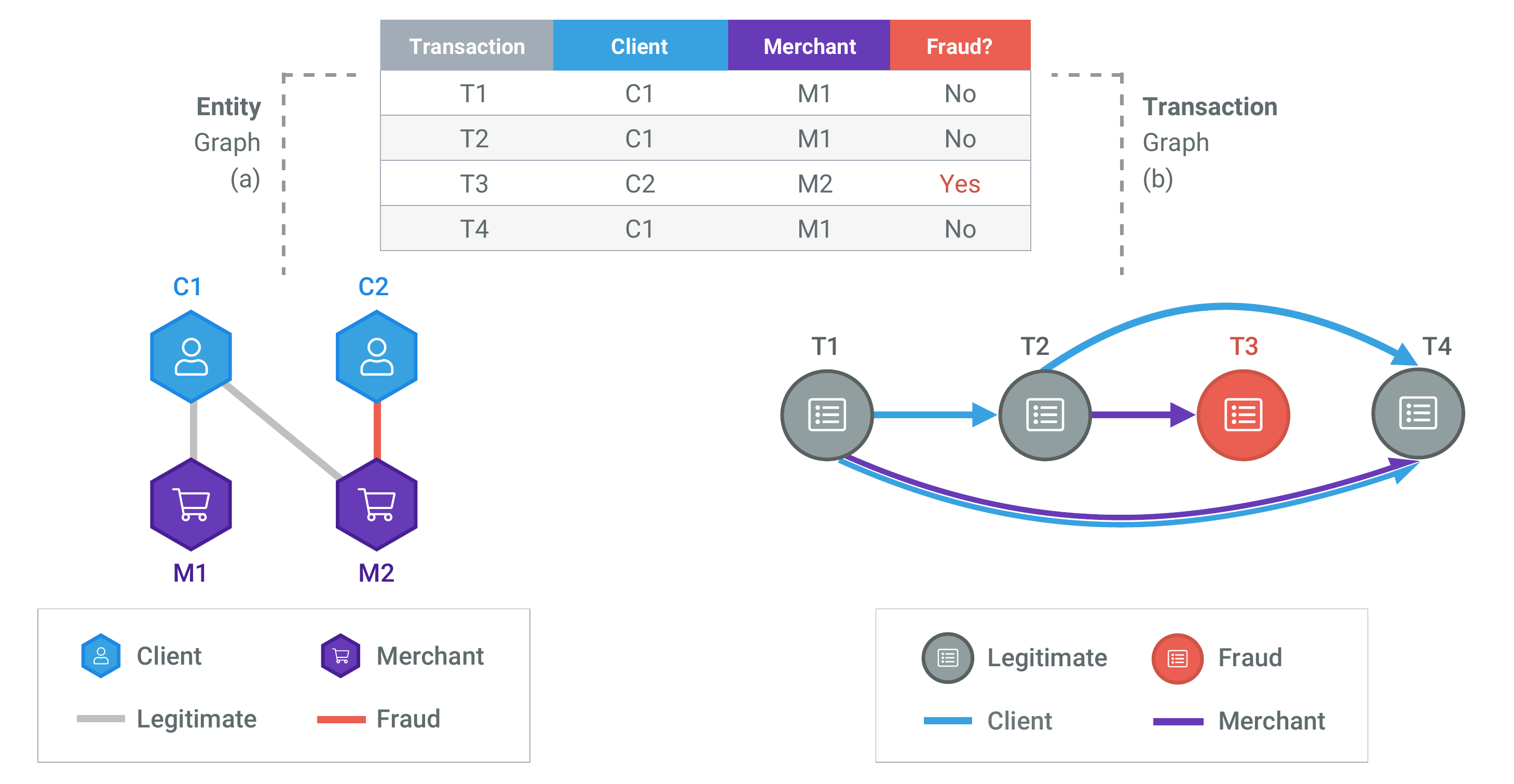}
	\caption{Example of (a) an entity graph and (b) a transaction graph.}
	\label{fig:ent_trx_graph_generation}
\end{figure*}

\section{Method}\label{sec:method}


In this section we present our methodological choices. First, we discuss how we build networks from banking datasets (Section~\ref{sec:method_graph_representation}). Then, we introduce heterogeneous network motifs and describe how to compute and identify these motifs in banking datasets, with a special focus on the null model and the measure of motif significance.  (Section~\ref{sec:method_network_motifs})


\subsection{Graph representation}\label{sec:method_graph_representation}

Banking datasets usually consist of transactions between entities, such as people, merchants, and businesses. They can include different types of transactions such as card payments or bank transfers. 

From a banking dataset, we can build two graph representations, namely (a) entity graphs and (b) transaction graphs, as illustrated in  Figure~\ref{fig:ent_trx_graph_generation}.

\subsubsection{Entity graph}\label{sec:graph_representation_ent_graph}

In an entity graph, $G = (V, E)$, nodes $V$ represent entities, such as merchants or clients, and edges $E$ connect entities with at least one shared transaction. This way of representing banking datasets is helpful to highlight suspicious entity behavior.

Consider, for example, the entity graph in Figure~\ref{fig:ent_trx_graph_generation} (a). There, we represent C1 and M1 as two connected nodes of different types, i.e., $\{C1, M1\} \in V$, $(C1, M1) \in E$, and $\phi(C1) \neq \phi(M1)$, where $\phi(u)$ is the type of node $u$. Since we connect all $k$ entities in a transaction, they form a $k$-node clique in the graph. When the same two entities are parties to multiple transactions (e.g., a person makes several purchases at a retail store), we aggregate the transaction information into a single edge $(u, v) \in E$.

Therefore, an entity graph is undirected and heterogeneous. The node label $\phi(u)$ corresponds to the entity's type (i.e., client, card, merchant, or terminal). The edge label $\mu(u, v)$ is binary, indicating whether there is at least one fraudulent transaction involving the two entities.

\subsubsection{Transaction graph}\label{sec:graph_representation_trx_graph}

In a transaction graph, $G = (V, E)$, nodes $V$ represent individual transactions, and edges $E$ connect transactions that share entities. 

In the transaction graph from Figure~\ref{fig:ent_trx_graph_generation} (b), transactions T1 and T2 are represented as two nodes connected by an edge indicating that the same client made both, i.e., $\{T1, T2\} \in V$ and $(T1, T2) \in E$. Since connecting all transactions with common entities would result in very dense graphs, we only connect transactions that occurred within a time window, e.g., transactions made by a client in less than 24 hours. Moreover, we use edge direction to encode the temporal sequence of transactions, with edges connecting older transactions to more recent ones, i.e., if $(u, v) \in E$, then $v$ is more recent than $u$. If two transactions occur in the same timestamp, we add a bidirectional edge between the two nodes, i.e., $\{(u,v), (v,u)\} \in E$.

Thus, a transaction graph is directed and heterogeneous. The node label is binary (the transaction is fraudulent or legitimate), i.e., $\phi(x) \in \{fraud, legit\}$. The edge label is one of $2^k-1$ possible labels, all the combinations of sharing one or more of $k$ different entities, i..e, $\mu(u, v) \in M, |M| = 2^k-1$. In Figure~\ref{fig:ent_trx_graph_generation}, $k = 2$, hence there are 3 possible labels for edges, $M = 3$.

Transaction can be used to investigate patterns between transactions. Additionally, machine learning classification models can benefit from receiving topological information (including centrality measures or node embeddings) extracted from transaction graphs \cite{oliveira2021guiltywalker}.

\begin{figure*}[!t]
	\centering
	\includegraphics[width=0.9\textwidth]{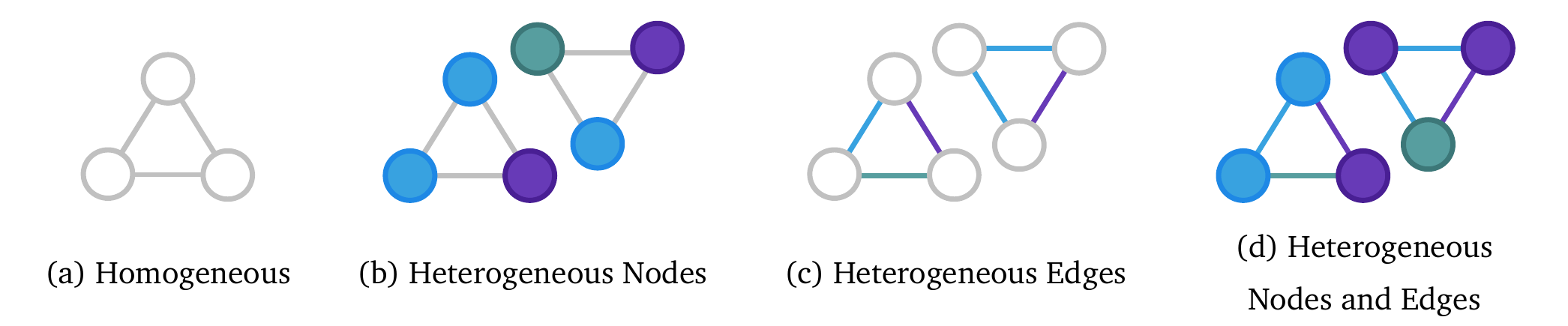}
	\caption{Different 3-node cliques. In our case, we use both heterogeneous nodes and edges.}
	\label{fig:3nodesubgraph}
\end{figure*}

\subsection{Heterogeneous network motifs}\label{sec:method_network_motifs}

A \emph{network motif} is a subgraph that appears \emph{more frequently than expected}. The concept of appearing more frequently than expected commonly relies on building a large set of randomized networks $\mathcal{R}_G$ that are \emph{similar} to the original network~\cite{milo2002}. In the literature, authors typically use either 100 or 1000 randomized networks~\cite{ribeiro2009strategies}.

Suppose the frequency of a given subgraph $m_i$ in the original network is, according to some significance measure (Section~\ref{method_network_motifs_motif_significance}), \emph{much higher} than the (average) frequency on similar randomized networks, i.e., $freq(m_i, G) >> average\_freq(m_i, \mathcal{R}_G)$. In that case, subgraph $m_i$ is a network motif. Similarly, if the subgraph's frequency is \emph{much lower} than the (average) frequency on similar randomized networks, that subgraph is an anti-motif. In this work, we are interested in both motifs and anti-motifs.

Motif discovery involves computing the frequency of a given set of subgraphs and entails subgraph counting~\cite{ribeiro2019survey}. Subgraph counting receives as input a graph \textit{G} and a list of non-isomorphic subgraph types (e.g., all possible unique subgraphs with four nodes). Then, it outputs the frequencies of each subgraph type in \textit{G}, e.g., the frequencies of 4-node cliques, 4-node chains, or 4-node stars.

In this work, we use \emph{g-tries} for subgraph counting since they are a general framework able to count subgraphs of arbitrary size in heterogenous graphs~\cite{ribeiro2014g,ribeiro2014discovering}. Other approaches are faster for counting specific subgraphs, but, as far as we know, g-tries are the only method that support directed and heterogenous graphs~\cite{ribeiro2019survey}.

Since our graphs are heterogeneous, we need network motifs that consider node and edge heterogeneity. Concretely, we need to extract heterogeneous motifs and anti-motifs.

As an illustrative example, consider a 3-node clique where nodes and edges are homogeneous (Figure~\ref{fig:3nodesubgraph} (a)) and the following heterogeneous graph settings: (i) if nodes can be of two different types, there are four possible 3-node cliques (Figure~\ref{fig:3nodesubgraph} (b)), (ii) if edges can be of two different types, there are four possible 3-node cliques (Figure~\ref{fig:3nodesubgraph} (c)). Thus, in our work, disregarding node or edge labels undermines the necessary differentiation of topological structures. Heterogeneous motif discovery is more informative than traditional homogeneous motif discovery and more complex to extract and analyze.

In banking fraud analysis, heterogeneous network motifs are more helpful than homogeneous motifs. For instance, knowing that "clients connected to many different cards are more likely to be fraudulent" is arguably more informative than just knowing that "dense subgraphs can be indicative of fraud". 

\subsubsection{Network randomization}\label{method_network_motifs_null_model}

Computing the \emph{expected} frequency of a given subgraph requires randomizing the original network so that randomized networks are similar to the original. However, defining network \emph{similarity} is non-trivial and task-dependent. In practice, the following two approaches are common:
\begin{itemize}
	\item Initialize the randomized graph as a copy of the original graph and iteratively swap random pairs of edges~\cite{ribeiro2009strategies}. This method preserves vertices' in- and out-degrees.
	
	\item Initialize the randomized graph with the same nodes as the original graph and incrementally add edges with probabilities based on each nodes' degree in the original graph~\cite{milo2002}. This method provides an approximation of the vertices' in- and out-degrees.
	
\end{itemize}

However, these strategies are unsuitable to the banking fraud domain as they fundamentally change the semantic structure of banking graphs. For instance, in an entity graph, entities of the same type are never directly connected by an edge. However, adding or removing edges indiscriminately (while preserving each node's degree) will eventually lead to connecting nodes of the same type and thus compromise the validity of the randomized network. Since these strategies do not take node and edge labels into account, we need to follow a different approach for network randomization.

Instead of randomizing the original graph directly, we apply a randomization procedure directly on the tabular data and then build the randomized networks from the randomized tabular data. Thus, the randomization procedure works as follows: 

\begin{enumerate}
	\item Shuffle all the fraud labels. This step maintains the fraud rate of the original dataset but randomizes its attribution to different transactions.
	\item Iterate over each entity type and, for each entity type, we randomly chose $\rho *m$ pairs of values to be swapped. Here, $\rho \in [0, 1]$ controls how much we want to randomize the original network, and $m$ is the number of transactions.
	\item Build the resulting graph from the randomized data, as described in Sections~\ref{sec:graph_representation_ent_graph} and ~\ref{sec:graph_representation_trx_graph}.
\end{enumerate}

This network randomization strategy guarantees semantically valid banking networks with a random topology. 

\subsubsection{Motif significance measures}\label{method_network_motifs_motif_significance}

After doing subgraph counting on both the original network and the randomized networks, we use a motif significance measure to evaluate which subgraphs are over- and under-represented. Several measures have been proposed~\cite{xia2019}, but here we focus on two of the most well-established: the \emph{z-score} and the \emph{ratio}.

The z-score of a subgraph, $z_i$, is computed as follows:

\begin{equation}
	z_i = \frac{ f_i^{o} - \mu_i^{R} }{ \sigma_i^{R} }
\end{equation}

Where,

\begin{itemize}
	\item $f_i^{o}$ is the frequency of subgraph $i$ in the original network.
	\item $\mu_i^{R}$ and $\sigma_i^{R}$ are the mean and standard deviation of the frequencies of subgraph $i$ in the randomized networks, respectively.
\end{itemize}

We compute the ratio of a subgraph, $r_i$, as follows:

\begin{equation}\label{eq:ratio}
	r_i = \frac{ f_i^{o} }{ \mu_i^{R} }
\end{equation}

Our goal is to find the subgraphs with the highest z-scores/ratio (i.e., the motifs) and subgraphs with the lowest z-scores/ratio (i.e., the anti-motifs).

In our experiments (Section~\ref{sec:results}), we show the ratio of the subgraphs since it is more interpretable than the z-score: if $r_i = 100$, the subgraph appears 100 times more often in the original network than in the randomized networks. We complement our analysis by reporting $f_i^{o}$, $\mu_i^{R}$, and $\sigma_i^{R}$.

\section{Results}
\label{sec:results}


\subsection{Data overview}
\label{subsec:data}

Table~\ref{tab:schema_network} outlines the parameters used to generate the entity graph and the transaction graph and provides summary statistics on the two graphs. 

In the entity graph, we consider four node types (i.e., client, merchant, terminal, and card) and two edge types (i.e., fraudulent or legitimate). 

In the transaction graph, we consider two node types (i.e., fraudulent and legitimate) and three edge types (i.e., only client, only merchant, or both client and merchant). The rationale for choosing these two entities lies in the close relationship between clients and cards and, similarly, merchants and terminals. In other words, since most clients use few cards and most merchants have few terminals, we simplify the graph by dropping the card and terminal entities.

Additionally, we found it necessary to constrain the considered time window due to computational constraints on subgraph counting.

Both graphs have hundreds of thousands of edges and many different connected components. The fraud rate in the entity graph is the number of fraudulent edges, while, in the transaction graph, the fraud rate is the number of fraudulent nodes. As a result, the fraud rates differ depending on the graph type.

\begin{table}[!b]
	\caption{Parameters used for graph generation and general statistics (NA means Not Applicable).}
	\small
	\centering
	\begin{tabular}{@{}lcc@{}}
		\toprule
		\multicolumn{1}{c}{}    & entity graph      & transaction graph   \\ \midrule
		period           & 2 days, 5 hours    & 1 day, 21 hours           \\
		lookback period         & NA                 & 6 hours                   \\
		random networks   & \multicolumn{2}{c}{100}                        \\
		$\rho$    & \multicolumn{2}{c}{0.8}                        \\ \midrule
		nodes                & 51,428              & 20,209                     \\
		edges                & 104,833             & 214,054                    \\
		node types           & 4                  & 2                         \\
		edge types           & 2                  & 3                         \\
		conn. components & 5,796               & 9,877                      \\
		fraud rate              & 1:1000         & 1:500               \\ \bottomrule
	\end{tabular}
	\label{tab:schema_network}
\end{table}

\subsection{Motif analysis}
\label{subsec:motif_analysis}

In this subsection, we start with the entity graph results and then show the results of the transaction graph.

We analyze which subgraphs are motifs and anti-motifs based on the ratio (Equation~\ref{eq:ratio}): we consider subgraphs that have a considerably higher ratio than the others to be motifs, and subgraphs with the lowest ratios to be anti-motifs.

We focus on subgraphs of size three due to the higher computational cost of larger subgraphs.

\subsubsection{Entity graph}
\label{subsubsub:ent-ent}

\begin{figure}[!t]
	\centering
	\includegraphics[width=0.49\textwidth]{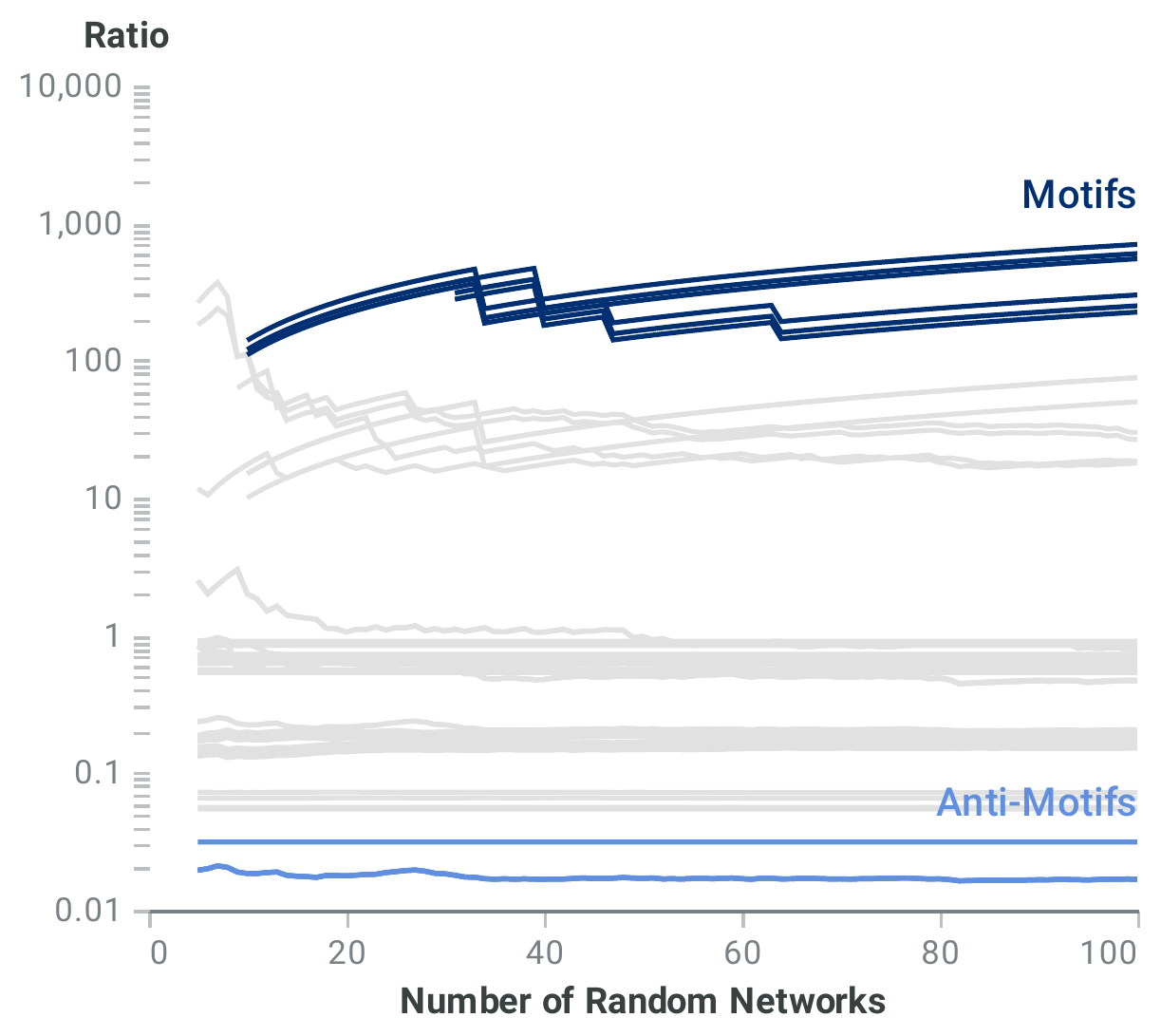}
	\caption{Entity graph's ratio evolution, highlighting the top-6 motifs in dark blue and the top-4 anti-motifs in light blue.}
	\label{fig:subgs_ee_ratio_evol}
\end{figure}

\begin{figure}[!t]
	\centering
	\includegraphics[width=0.48\textwidth]{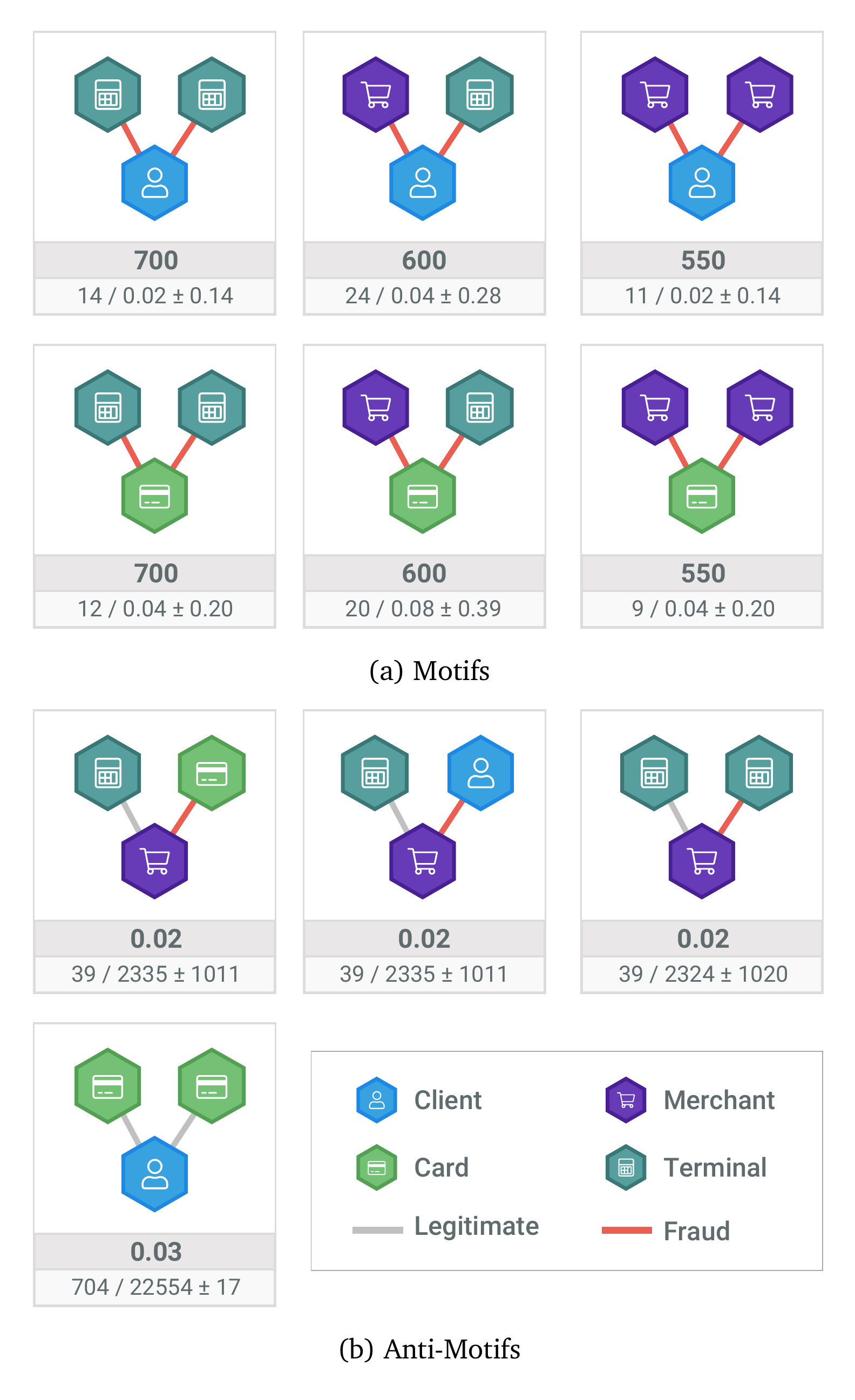}
	\caption{Entity graph motifs and anti-motifs. For each subgraph $i$, we show its ratio $r_i$ (first line) and $f^o_i$ / $\mu^R_i \pm \sigma^R_i$ (second line).}
	\label{fig:subgs_ee_ratio_top}
\end{figure}

From Figure~\ref{fig:subgs_ee_ratio_evol} we can see that, for most cases, the ratio of all subgraphs stabilizes relatively quickly at $\approx$ 40 random networks. We observe that a few subgraphs have significantly higher ratio than the others ($r_i > 100$).  Moreover, a few subgraphs have significantly lower ratio than the others ($r_i\approx 0.01$). We consider the first to be the motifs, and the last to be the anti-motifs, shown in Figure~\ref{fig:subgs_ee_ratio_top} (a) and Figure~\ref{fig:subgs_ee_ratio_top} (b), respectively.

From Figure~\ref{fig:subgs_ee_ratio_top} we notice that none of the subgraphs is a 3-node clique. This result might seem counter-intuitive as all transactions form a 4-node clique. A possible explanation is that some merchants are hub nodes and thus induce chain-like subgraphs. Another general observation is that all edges in the motifs are fraudulent, which is not true for the anti-motifs. Domain knowledge indicates that fraudulent transactions tend to be closely connected. 

The first three motifs from Figure~\ref{fig:subgs_ee_ratio_top} (a) have a client at its center, connected to either (i) two terminals, (ii) one terminal and a merchant, or (iii) two merchants. Subgraphs (i) and (iii) tell us that the client made two fraudulent transactions at different merchants and terminals, respectively. These two subgraphs are motifs because fraud is sometimes recurring for fraudulent clients, and this information is lost in the randomized networks since they reshuffle fraud labels. Subgraph (ii) tells us a similar story but notice that, since the merchant and the terminal are not connected, the merchant and the terminal are from different transactions. The last three motifs are equivalent to the first three but with the card at the center of the subgraph. Since clients can use multiple cards, the original network counts are lower for subgraphs with the card at the center of the chain than the counts of the subgraphs with the client at the center of the chain. We also note that, in practice, it might be interesting to investigate all cards used by the client.

The first three anti-motifs from Figure~\ref{fig:subgs_ee_ratio_top} (b) have a merchant at its center. All three anti-motifs convey the same information: merchants typically either have only legitimate transactions or only fraud transactions, i.e., fraud tends to cluster around the same risky merchants. In the randomized networks, since we randomly swap edges, these relations are lost. Finally, the last anti-motif tells us that it is not common for clients to use multiple cards in legitimate transactions. Indeed, clients that use multiple cards are typically associated with fraudulent activity.

\subsubsection{Transaction graph}

\begin{figure}[!t]
	\centering
	\includegraphics[width=0.49\textwidth]{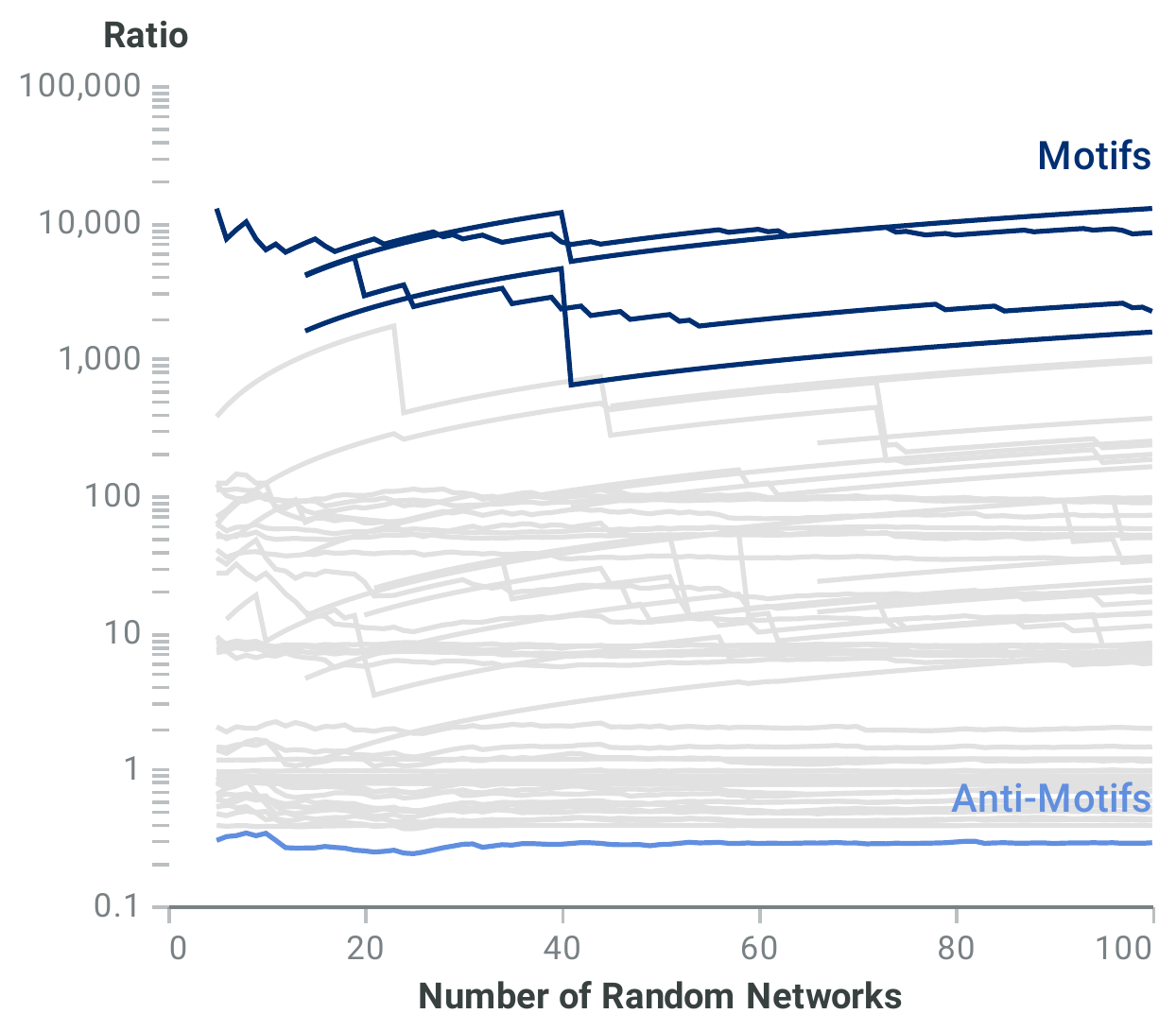}
	\caption{Transaction graph's ratio evolution, highlighting the top-4 motifs in dark blue and the top-1 anti-motif in light blue.}
	\label{fig:subgs_tt_ratio_evol}
\end{figure}

\begin{figure}[!t]
	\centering
	\includegraphics[width=0.48\textwidth]{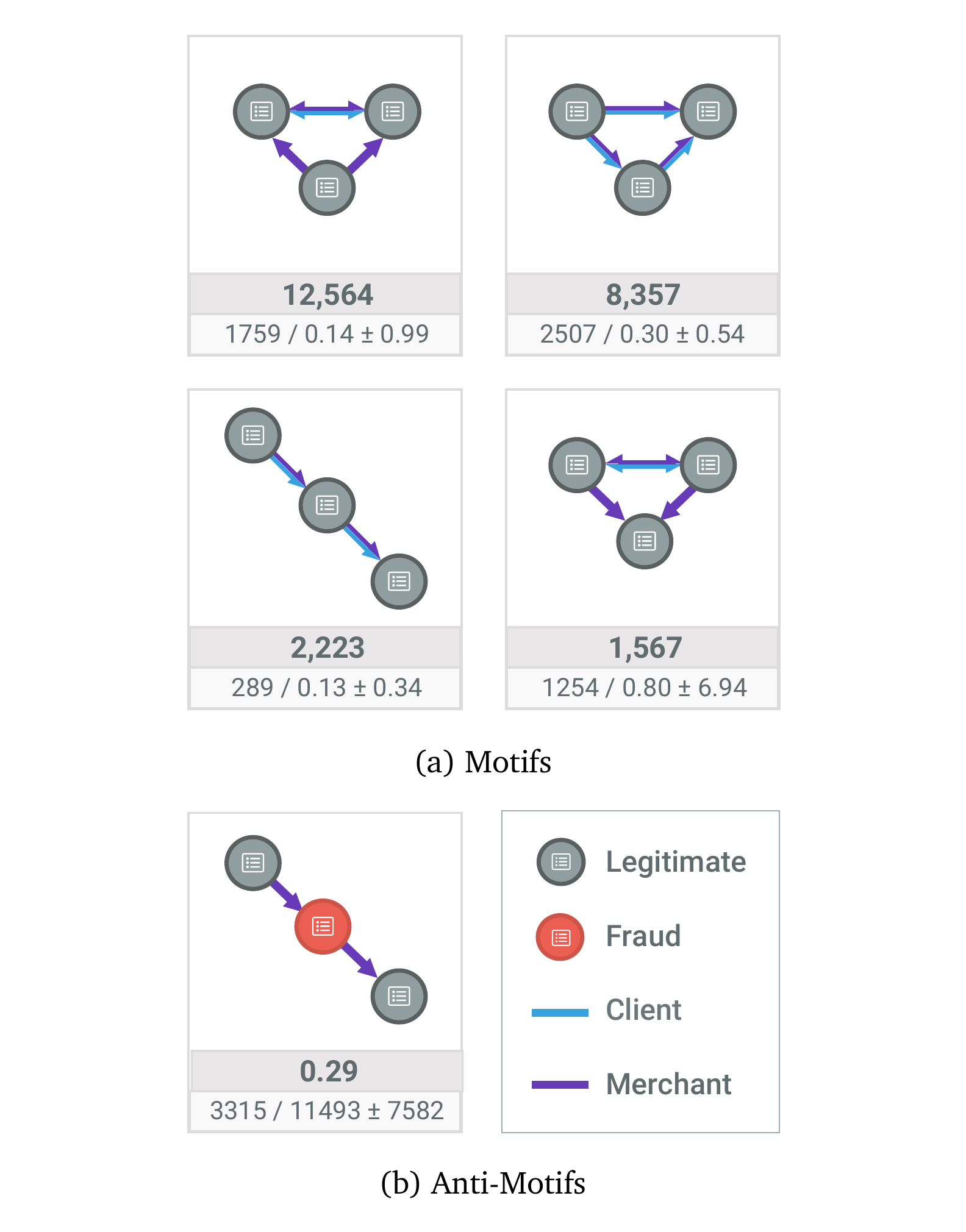}
	\caption{Transaction graph motifs and anti-motifs.  For each subgraph $i$, we show its ratio $r_i$ (fist line) and $f^o_i$ / $\mu^R_i \pm \sigma^R_i$ (second line).}
	\label{fig:subgs_tt_ratio}
\end{figure}

Figure~\ref{fig:subgs_tt_ratio_evol} shows the evolution of the ratio for each 3-node subgraph in the transaction graph. After 80 random networks, the ratios seem to stabilize, and we can clearly distinguish five subgraphs that stand out, namely, the top-4 subgraphs and the bottom-1 subgraph. These subgraphs are presented in Figure~\ref{fig:subgs_tt_ratio}.

The common feature in the four motifs is at least two transactions that share the same client and merchant. We may lose this pattern during the randomization since we swap merchants and clients independently. However, it is still interesting to observe a significant number of patterns where the same client makes two or three transactions in the same merchant in less than six hours.

Three of the four top motifs are a 3-node clique, which happens when three transactions that share a merchant or client occur within the same six-hour window. Once again, the randomization of the entities breaks such patterns.

It is important to note that some motifs contain transactions processed in the same timestamp (represented with bi-directional edges). This pattern can happen in the same merchant when the merchant is processing transactions in a batch or has multiple terminals.

The only anti-motif is a sequence of three transactions in the same merchant, where the middle transaction is fraudulent, and the other two are legitimate. This pattern is less frequent in the original network than in the randomized networks since fraud transactions typically occur together in reality, and the randomized networks break this pattern.

\section{Usage Scenarios}
\label{sec:usage}

The motif analysis can be a tool to characterize banking datasets. Beyond summary statistics, the list of motifs and anti-motifs surfaces underlying fraud patterns. 

Fraud experts, who may not be knowledgeable in data science or statistics, often use graph visualization for data exploration. Motif analysis serves as a visual summary characterization of a dataset for fraud detection.

As an illustrative example, let us consider a fraud expert at a bank. Their role entails designing new rules to prevent fraud and reviewing unlabelled transactions. The expert can review the characteristic patterns surfaced by motif analysis to tailor the fraud detection system in place. Over time, fraudsters design new schemes to evade  detection. Periodic analysis of motifs surfaces upcoming fraud schemes and overall behavioral trends.

When reviewing an unlabelled transaction, the expert can compare the respective subgraph with known motifs. This context can give insight into which pattern, fraudulent or legitimate, might occur.

On the other hand, by extracting the most relevant transaction patterns and uncovering new fraud schemes, motif analysis can complement common fraud detection systems. These insights can be used to improve both machine learning models and rule-based systems.

\section{Conclusions}
\label{sec:conclusion}

We explore motifs and anti-motifs in the context of banking fraud using two graph representations, namely entity graphs and transaction graphs. We propose a novel randomization method that operates directly on tabular data. This way, we overcome the limitations of current network randomization methods in the context of banking fraud.

Moreover, we extract heterogeneous network motifs that convey more information than traditional network motifs and find they offer interpretable results. Insights extracted from motif analysis can be used to aid fraud analyst investigate specific cases and improve fraud detection systems by uncovering new fraud schemes.


As future work, one can investigate whether different banking datasets have similar motifs (and anti-motifs) and if those patterns are different in merchant datasets. This research would follow the findings by \citet{milo2004superfamilies} where they report that networks with similar contexts have similar subgraph patterns. One can also extend the analysis to larger motif sizes, different temporal windows, and the inclusion of transaction amounts or fraud labels as graph properties.


\bibliographystyle{ACM-Reference-Format}
\bibliography{references}

\end{document}